\documentstyle[epsfig,12pt]{article}
\begin{document}
\begin{titlepage}
\vskip1in
\begin{center}
{\Large {\bf Reflection Factors for the Principal Chiral Model}}
\end{center}
\vskip1in
\begin{center}
{\large
J.N.Prata

Department of Mathematical Sciences

University of Durham

South Road

Durham, DH1 3LE, England}

{\it j.n.g.n.prata@durham.ac.uk}
\end{center}
\vskip1in
\begin{abstract}
We consider the $SU(2)$ Principal Chiral Model (at level $k=1$) on the 
half-line with scale invariant boundary conditions. By looking at the IR 
limiting conformal field theory and comparing with the Kondo problem, we 
propose the set of permissible boundary conditions and the corresponding 
reflection factors.
\end{abstract}
\end{titlepage}

\section{Introduction}

The principal chiral model $PCM_k$ is defined by the action:
\begin{equation}
S_{PCM_k} = \frac{1}{2 \lambda^2} \int_{\partial {\cal B}} Tr
\left\{(g^{-1} \partial_{\mu} g) ( g^{-1} \partial^{\mu} g) \right\}
d^2x + i k \Gamma (g),
\end{equation}
where $\Gamma(g)$ is the Wess-Zumino-Witten (WZW) term \cite{Witten}:
\begin{equation}
\Gamma (g) = \frac{1}{24 \pi} \int_{{\cal B}} Tr \left\{(g^{-1}
  \partial_{\mu} g) (g^{-1} \partial_{\nu} g)(g^{-1} \partial_{\lambda}
  g) \right\} \epsilon^{\mu \nu \lambda} d^3 x.
\end{equation}
In eq.(1) $g$ is a Lie group valued field defined on a two-dimensional
compact spacetime surface $\partial {\cal B}$. The region of
integration ${\cal B}$ in eq.(2) is a three-dimensional
simply-connected manifold whose boundary is $\partial {\cal
  B}$. Topological arguments show that the ambiguity in this definition
amounts to the WZW functional (2) being determined up to a positive
integer, which can be reabsorbed into the constant k in eq.(1)
\cite{Novikov}. If the Lie group is simple this ensures the 
positivity of the action (1). Here we shall consider the group $G=SU(2)$. The 
action (1) hence enjoys a $SU(2)_L \times SU(2)_R$ global symmetry.

For $k=0$ the theory corresponds to a $O (4) \approx SU(2) \times SU(2)$ 
nonlinear sigma model. Its behaviour is massive. If $k \ne 0$ the 
renormalization group (RG) analysis reveals that it interpolates between 
two fixed points \cite{Zam1}. The ultraviolet (UV) fixed point is 
controlled essentially by the first term in eq.(1). The RG flow of the 
coupling $\lambda^2$ terminates at the infrared (IR) fixed point 
$\lambda^2 = 8 \pi /k$ where the theory becomes massless at all 
distances. At this point the theory is characterized by a conformal 
field theory based on two $SU(2)_k$ Kac-Moody algebras (at level $k$). 
For a generic k the RG trajectory arrives at the IR fixed point along 
the direction defined by the irrelevant field $Tr (g^{-1} \bar{\partial} 
g g^{-1} \partial g)$ of dimension $1+2/(k+2)$ \cite{Knizhnik}. For 
$k=1$, this field does not exist in the conformal theory and the 
incoming direction is defined by the operator $T \bar T$, composed from 
the components of the stress tensor of the IR conformal field theory.
The point where the model crosses over from the region of one fixed
point to the other introduces a mass scale that breaks the scale
invariance.

 The $PCM_k$ was argued to be integrable in refs.
\cite{Abdalla}-\cite{Polyakov1} and its thermodynamic 
Bethe Ansatz (TBA) equations proposed in \cite{Zam2}. Zamolodchikov's 
c-function was shown to take the values $c_{UV} = 3$ and 
$c_{IR} = 3k/(k+2)$ at the fixed points. The latter is in agreement 
with the central charge of the $SU(2)_k$ conformal 
field theory \cite{Witten}, \cite{Knizhnik}. Al.B.Zamolodchikov 
and A.B.Zamolodchikov subsequently proposed the background scattering 
in terms of massless particles that leads to the correct TBA 
equations for $k=1$ \cite{Zam1}. Following a prescription developed by 
Smirnov and Kirillov \cite{Kirillov} in the context of the 
$SU(2)$-invariant Thirring model, they also showed that the form factors 
associated with the chiral currents obey the correct commutation 
relations. However the central term remains undetermined. 
Notwithstanding this, it can be shown to take the correct value by TBA 
analysis, \cite{Zam1}. Mejean and Smirnov \cite{Mejean} derived the form 
factors for the trace of the stress tensor.

In this work, we study the model in the presence of reflecting
boundaries. This problem may appear awkward at the first sight due to
the very definition of the action (1). This is because the boundary of
a boundary is evidently an empty set. However we shall circumvent this
obstacle by ignoring the classical action (1) altogether and going
directly to the quantum theory. The drawback of this approach is that we are 
not able to apply useful information from the classical theory, 
\cite{Moriconi1}, \cite{Moriconi2}, \cite{Bowcock}. The determination of the 
boundary conditions compatible with integrability and the corresponding
reflection amplitudes will involve some amount of guesswork. We shall
use as guideline some knowledge coming from the symmetries of the
problem, the limiting IR conformal field theory and a related problem
(Kondo). The difference between this and the Kondo problem lies
in the fact that in the former the scale invariance is broken in the
bulk by a mass scale associated with a very unstable $O(4)$-isovector
resonance \footnote{We assume that the boundary introduces no 
additional mass scale.}, whereas in the Kondo problem the scale
invariance is broken at the boundary.

Let us first assemble some known results about this model, \cite{Zam1}. 
The spectrum of the theory consists of stable massless particles: 
left-movers and right-movers. It is convenient to parametrise the 
on-mass-shell momenta of the particles in terms of the rapidity 
variables $- \infty < \beta, \beta' < \infty$:
\begin{equation}
\left\{
\begin{array}{l l l}
e= p & = \frac{M}{2} e^{\beta} & , \mbox{for right-movers},\\
& & \\
e= - p & = \frac{M}{2} e^{- \beta'} & ,\mbox{for left-movers}.
\end{array}
\right.
\end{equation}

With this parametrisation opposite momenta still correspond to opposite
rapidities, \cite{Fendley3}. For left-left and right-right scattering 
all Mandelstam variables vanish and since the scattering depends only on 
the dimensionless ratios of the momenta, the mass scale $M$ is
arbitrary. The right-left scattering on the other hand distinguishes
some preferable scale normalization $M$. The Mandelstam variable is
now:
\begin{equation}
s= M^2 exp (\beta_1 - \beta_2),
\end{equation}
for the scattering of a right-mover with rapidity $\beta_1$ and a left-mover with
rapidity $\beta_2$. The soft scattering corresponds not to $\beta_1 \sim
\beta_2$ but rather to $\beta_1 - \beta_2 \to - \infty$. The mass scale
$M$ can thus be chosen so that the crossover occurs at $\beta_1 \sim
\beta_2$.

Besides being massless these particles form doublets $(u,d)$ under the 
global $SU(2)$ symmetry. However there is an additional structure: each
particle is also a kink, \cite{Fendley1}. The $SU(2)_k$ WZW model has 
$(k+1)$ degenerate vacua. The allowed kinks interpolate between adjacent 
vacua. So, for instance, each left-moving particle doublet can be 
labeled by $(u_{c,c  \pm 1}^{(L)}, d_{c, c \pm1}^{(L)} )$, where $c$ is 
an index referring to the vacua $(c=1,2, \cdots , k+1)$. In the simplest 
case $(k=1)$, which is the one we are interested in, the only 
nontrivial structure is that of a $(u,d)$ doublet. We then represent 
the $SU(2)_R$ doublet of right-moving particles by the symbol 
$R_a (\beta)$ ($a= \pm$ is the right isotopic index) and the $SU(2)_L$ 
doublet of left-movers by $L_{\bar a} (\beta')$ ($\bar a = \pm$ is the 
left isotopic index) with energy spectra given by eq.(3).

The charge conjugation operator $C$ is defined with respect to the
$SU(2)$ symmetry by:
\begin{equation}
C= \left(
\begin{array}{r l}
0 & 1\\
-1 & 0
\end{array}
\right).
\end{equation}
We shall denote the antiparticles of $R_a (\beta)$ and $L_{\bar a}
(\beta')$ by $\bar R_{\underline a} (\beta)$ and $\bar
L_{\bar{\underline a}} (\beta')$, respectively. Let us now consider the
general $2 \to 2$ scattering of a particle $R_a(\beta_1)$ with its
antiparticle $\bar R_{\underline b} (\beta_2)$. The S-matrix element is
given by \cite{Berg}:
\begin{equation}
\begin{array}{c}
^{out} <R_c (\beta_1') \bar R_{\underline d} ( \beta_2') | R_a
(\beta_1) \bar R_{\underline b} ( \beta_2) >^{in} =  \delta (\beta_1' - 
\beta_1) \delta (\beta_2' - \beta_2) F_{ab}^{cd} (\beta) \\
\\
- \delta (\beta_1' - \beta_2) \delta (\beta_2' - \beta_1) B_{ab}^{dc} (\beta),
\end{array}
\end{equation}
where $\beta \equiv \beta_1 - \beta_2$. The forward and backward
amplitudes are:
\begin{equation}
\left\{
\begin{array}{l}
F_{ab}^{cd} (\beta) = \delta_a^c \delta _b^d u_1 (\beta) + \delta_{ab}
\delta^{cd} v_1 (\beta),\\
\\
B_{ab}^{dc} (\beta) = \delta_a^c \delta _b^d u_2 (\beta) + \delta_{ab}
\delta^{dc} v_2 (\beta).
\end{array} \right.
\end{equation}
However, for massless particles backward scattering is unacceptable and
we therefore set $u_2 (\beta) = v_2 (\beta) = 0$.

The particle-particle S-matrix element is given by:
\begin{equation}
R_a (\beta_1) R_b (\beta_2) = S^{cd}_{ab} (\beta_1 - \beta_2) R_d (\beta_2) 
R_c (\beta_1),
\end{equation}
with
\begin{equation}
S_{ab}^{cd} (\beta) = \sigma_T (\beta) \delta_a^c \delta_b^d + \sigma_R
(\beta) \delta_a^d \delta_b^c.
\end{equation}
$\sigma_T (\beta)$ and $\sigma_R (\beta)$ are the transition and
reflection amplitudes, respectively. It is also convenient to introduce
the 2-particle amplitude in the isovector and isoscalar channels:
\begin{equation}
\left\{
\begin{array}{l}
S_V (\beta) = \sigma_T ( \beta) + \sigma_R (\beta),\\
\\
S_0 ( \beta) = \sigma_T (\beta) - \sigma_R (\beta).
\end{array}
\right.
\end{equation}
Using the requirements of factorizability, unitarity and crossing
symmetry, the following minimal solution was suggested in
ref.\cite{Zam1}:
\begin{equation}
\left\{
\begin{array}{l}
u_1(\beta) = - \sigma_T (\beta) - \sigma_R ( \beta),\\
\\
 v_1 (\beta)
= \sigma_R (\beta),\\
\\
\sigma_T ( \beta) = \frac{i}{\pi} \beta \sigma_R ( \beta), \\
\\
\sigma_R (\beta) = - \frac{i \pi}{ \beta - i \pi} S_V ( \beta),
\end{array}
\right.
\end{equation}
where
\begin{equation}
S_V ( \beta) = \frac{ \Gamma \left( \frac{1}{2} + \frac{\beta}{2i \pi}
  \right) \Gamma \left( - \frac{\beta}{2 i \pi} \right)}{\Gamma \left(
  \frac{1}{2} - \frac{\beta}{2 i \pi} \right) \Gamma \left(
  \frac{\beta}{2 i \pi} \right) }.
\end{equation}
Of course we get exactly the same expression for the L-L
scattering. The non-trivial right-left scattering is defined by the
commutation relations:
\begin{equation}
R_a (\beta) L_{\bar a} ( \beta') = U_{a \bar a}^{b \bar b} ( \beta -
\beta') L_{\bar b} (\beta') R_b (\beta).
\end{equation}
As we discussed before, this scattering breaks the scale invariance
thus spoiling the $\hat{SU} (2) \times \hat{SU} (2)$ current algebra
symmetry. However action (1) is invariant under the global $SU(2)_L
\times SU(2)_R$ isotopic symmetry at all distances. The only form of
$U_{a \bar a}^{b \bar b}$ preserving this symmetry is:
\begin{equation}
U_{a \bar a}^{b \bar b} ( \beta) = U_{RL} ( \beta) \delta_a^b
\delta_{\bar a}^{\bar b}.
\end{equation}
The factorization constraint is trivially met for this choice. For
massless particles there is a combined unitarity-crossing restriction, 
\cite{Zam6}:
\begin{equation}
U_{RL} (\beta + i \pi) U_{RL} ( \beta) = 1.
\end{equation}
The simplest non-trivial solution proposed in ref.\cite{Zam1}
is\footnote{This solution has also the virtue of yielding a kernel of
the form $- i \partial/ \partial \beta log U_{RL} (\beta) = 1/ cosh \beta$, 
which coincides with that of the TBA equations for the magnons, \cite{Zam1}, 
\cite{Zam4}, \cite{Zam2}.}:
\begin{equation}
U_{RL} (\beta) = \frac{1}{U_{LR} (- \beta)} = tanh \left(
  \frac{\beta}{2} - \frac{i \pi}{4} \right).
\end{equation}
It is worth noting that both amplitudes (12) and (16) have no poles on
the physical sheet. Also, we see from the soft scattering $(\beta
\to - \infty)$ in eq.(16) that the fields behave as fermions. And since
$S_{aa}^{aa} (0) = 1$, we conclude that we have a selection rule
preventing any two particles of the same type to be in exactly the same
quantum state.

\section{Boundary interactions and the Kondo problem}

Let us assume that our system is now restricted to the truncated line
$x<0$ with a boundary located at $x=0$. We can interpret the boundary
as an impenetrable particle sitting at the origin. Furthermore we
consider that the boundary conditions satisfy the following
prerequisites:

(1) the integrability is preserved,

(2) the boundary does not introduce any additional mass scale,

(3) at the IR fixed point conformal invariance is conserved.

The last statement can be made more precise. If $|B>$ is the boundary
state in the Hilbert space of the conformal field theory, then
\cite{Ishibashi}, \cite{Onogi}:
$$
(J_n^a + \bar J_{-n}^a ) |B> = 0, 
$$
where $J_n^a$ are the generators of the current algebra. If the Virasoro 
modes, $L_n$, are constructed according to Sugawara's 
prescription, then:
$$
(L_n - \bar L_{-n} ) |B> = 0.
$$
Cardy \cite{Cardy1} \cite{Cardy2}
argued that these boundary states are in one-to-one correspondence 
with the set of conformal blocks. This has the immediate consequence 
that, since the conformal towers are labeled by the isospin $l=0,1/2, 
\cdots, k/2$ \cite{Gepner} of the corresponding primary operator, so must the 
boundary states. The modular properties of the 
Kac-Moody characters $\chi_l^{(k)}$ are encoded in the matrix $S^{(k)}$,
\begin{equation}
\chi_l^{(k)} ( \tau) = \sum_{l'=0}^{k/2} S_l^{(k)l'}\chi_{l'}^{(k)} (
-1/ \tau), \qquad (l =0,1/2, \cdots, k/2),
\end{equation}
where:
$$
S^{(k)l'}_l = \sqrt{\frac{2}{k+2}} sin \left\{ \frac{\pi (2 l + 1) ( 2 l' +1)}
{k+2} \right\}.
$$
The partition function on an annulus of width $R$ with boundary conditions
$(l,p)$ on each side and the points along the periodic direction identified 
modulo $L$ is given by:
\begin{equation}
Z_{lp} (\tau) = \sum_{l' = 0}^{k/2} N_{lp}^{l'} \chi_{l'}^{(k)} (
\tau),
\end{equation}
where $\tau = L/R$.
$N_{lp}^{l'}$ are the structure constants of the fusion rules, which
can be obtained from the Verlinde formula \cite{Verlinde}:
\begin{equation}
\sum_{i=0}^{k/2} S_i^{(k)j} N_{pl}^i = \frac{S_p^{(k)j}
  S_l^{(k)j}}{S_0^{(k)j}},
\end{equation}
with $j,p,l=0,1/2,\cdots, k/2$. All these results hold because the
conformal field theory $\hat{SU} (2) \times \hat{SU} (2)$ is diagonal
in the sense that the partition function of the theory on the torus
takes the diagonal form\footnote{This is a consequence of $SU(2)$ being
a simply-connected group (cf.\cite{Gepner})}:
\begin{equation}
Z_{torus} ( \tau) = \sum_{l=0}^{k/2} \chi_l^{(k)} (\tau) \chi_l^{(k)}
(\tau).
\end{equation}
As $L$ tends to infinity the dominant contribution to the partition function 
(18) is that of the state with lowest energy given by:
\begin{equation}
E^0_{lp} = \frac{\pi}{R} \left( L^{(0)}_0 - \frac{c}{24} \right),
\end{equation}
where $L^{(0)}_0$ is the lowest eigenvalue of $L_0$ in the representations 
$l'$ for which $N^{l'}_{lp} \ne 0$.
In the case $k=1$, we have two integrable representations,
corresponding to isospins $l=0$ and $l=1/2$. They have been identified
with the identity operator and the fundamental field $g$ in the WZW
action, respectively \cite{Zam5}. The modular matrix $S^{(1)}$ is given 
by, \cite{Cappelli}:
\begin{equation}
S^{(1)} = \frac{1}{\sqrt 2} \left(
\begin{array}{l r}1 & 1\\
1 & -1
\end{array}
\right).
\end{equation}
From (18), (19) we get:
\begin{equation}
\left\{
\begin{array}{l}
Z_{00} (\tau) = Z_{\frac{1}{2} \frac{1}{2}} (\tau) = \chi_0^{(1)}
(\tau),\\
\\
Z_{0 \frac{1}{2}} (\tau) = Z_{\frac{1}{2} 0} (\tau) =
\chi_{\frac{1}{2}}^{(1)} (\tau).
\end{array}
\right.
\end{equation}
So we look for two distinct boundary conditions that lead to the two
above in the IR limit.
The $\hat{SU} (2) \times \hat{SU} (2)$ conformal field theory with a
boundary can also be seen as the IR limit of another integrable model
where this time the RG flow is controlled by the boundary
interaction. This situation is connected to the Kondo problem.

The Hamiltonian of the multi-channel Kondo problem is that of a k-tuple of 
free massless fermions $\psi_i$ ($i=1, \cdots , k$) antiferromagnetically
coupled to a fixed impurity of spin S in 3 dimensional Euclidean
space\footnote{For a review see ref.\cite{Affleck1} and references therein.}. 
The interaction term is of the form $\lambda \delta (x) \sum_{i=1}^k S_j
\psi_i^{\dagger} \sigma^j \psi_i$ where $\sigma^j$ 
are the Pauli matrices. A variety of arguments show that this 
problem can be described by a (1+1)-dimensional theory on the 
half line with the impurity sitting at the boundary. Moreover the 
RG flow interpolates between an unstable UV fixed point where the 
impurity is decoupled ($\lambda =0$) and
a strongly coupled IR one where the spin of the impurity is
`{\it screened}' $(\lambda = 2/(k+2))$. As before the crossover introduces a 
scale $T_K$ called the Kondo temperature. In the bulk the spectrum of 
particles is the same as for the $PCM_k$. However due to the bulk scale 
invariance, the R-L scattering must be trivial. For $k=1$, the L-L and R-R 
scattering are given by (9), (11) and (12). Actually in this description we 
are only considering the $SU(2)_k$ spin symmetry in $SU(2)_k \otimes SU(k)_2 
\otimes U(1)$ and discarding the additional charge $(U(1))$ and ``flavour'' 
$(SU(k)_2)$ symmetries. This is because the impurity only couples to the spin 
degrees of freedom. The particles are again $SU(2)$ doublets with a 
kink structure.

Several cases have to be distinguished, \cite{Fendley1}, \cite{Affleck1}. In 
the underscreened case ($k<2S$) one electron from each species binds to the 
impurity effectively reducing its spin to $q \equiv S- k/2$. The impurity 
particle is thus a member of a $(2 q +1)$-dimensional $SU(2)$ multiplet. In 
the exactly screened case ($k=2 S$) the electrons completely screen the 
impurity and it behaves like a $SU(2)$ singlet. Finally, in the overscreened 
case ($k>2S$) the boundary particle develops a kink structure.

For $k=1$, the overscreened case is not allowed, as this would imply
$S=0$, which means that there is no coupling to the impurity. The
exactly screened case corresponds to $q=0$ and hence the impurity
behaves effectively as a $SU(2)$ singlet. In this case the reflection 
amplitude was found to be, \cite{Fendley1}:
\begin{equation}
S_{RB} (\beta) = tanh \left(\frac{\beta}{2} - \frac{i \pi}{4} \right).
\end{equation}
The underscreened case
arises when $S>1/2$. If $S=1$, then $q=1/2$ and the impurity
is a $SU(2)$ doublet. There is also the possibility $S>1$, which we
shall not consider here.
In the underscreened case $S=1$ the boundary has an isotopic spin index
associated with it. The reflection matrix is given by eq.(9)\footnote{These
amplitudes are obtained by using the fact that, because of the bulk
scale invariance, one can extend the system to the whole line by
only considering, say, right-movers. The amplitude for the scattering between 
the left-mover and the impurity then satisfies the usual requirements of 
unitarity, crossing-symmetry, etc.}.

\begin{figure}
\begin{center}
\scalebox{0.4}{\includegraphics[60,250][500,500]{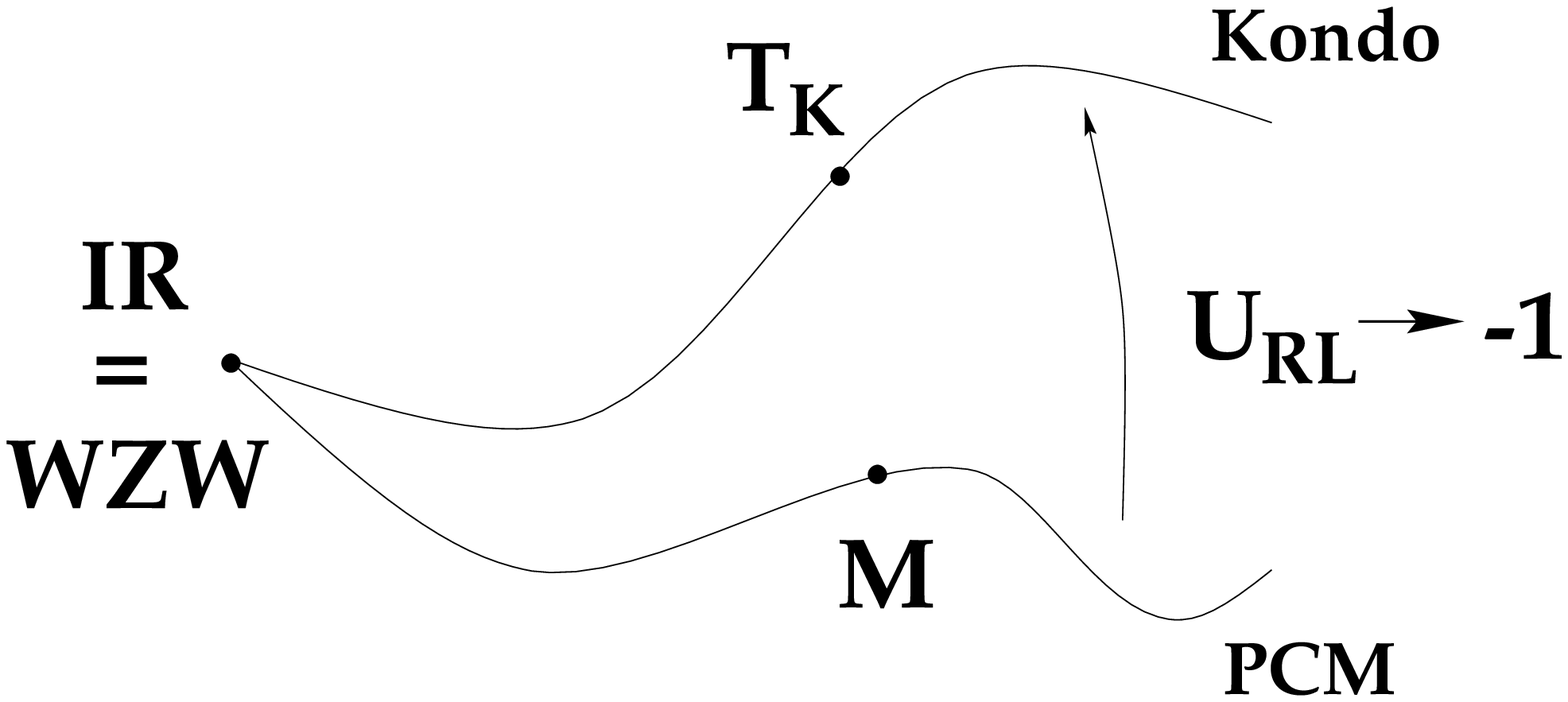}}
\caption{RG trajectories}
\label{fig:RG}
\end{center}
\end{figure}

Let us emphasize that our problem is not the Kondo
problem. Nevertheless, when we derive the boundary consistency 
equations, in the limit when the bulk theory becomes scale invariant,
the scattering amplitudes of the Kondo problem should solve these
equations. This will be an important consistency check. This program 
is depicted in fig.\ref{fig:RG}.

With these considerations in mind let us first admit that the boundary
impurity has effectively no spin. This would correspond to the exactly
screened case in the Kondo problem. We shall call this ``{\it fixed}'' 
boundary condition.

The reflection matrix $R_a^{\bar b}$ is defined by \cite{Ghoshal}:
\begin{equation}
R_a (\beta) B = \sum_{\bar b = \pm} R_a^{\bar b} (\beta) L_{\bar b} 
(- \beta) B.
\end{equation}
$B$ is formally an operator which, when acting on the vacuum, creates a 
boundary state $|B>$, i.e.
$$
|B> = B |0>.
$$
The fact that the boundary impurity is an $SU(2)$ singlet implies the
following diagonal form:
\begin{equation}
R_a^{\bar b} (\beta) = \delta_a^{\bar b} R_{RL} (\beta).
\end{equation}
The diagonal matrix (26) automatically satisfies the
boundary Yang-Baxter equation irrespective of $U_{RL}$. Let us now
consider the boundary crossing unitarity condition:
\begin{equation}
K^{\bar a b} (\beta) = \sum_{\bar c, d = \pm} K^{d \bar c} ( - \beta) 
U_{\bar c d}^{\bar a b} ( 2 \beta),
\end{equation}
where $K^{a \bar b} (\beta) = R_{\underline a}^{\bar b} ( i \pi /2 -
\beta) $. From eqs.(14) and (26) we get:
\begin{equation}
R_{LR} ( - \beta) = - \frac{R_{RL} ( i \pi + \beta)}{U_{RL} ( 2 \beta)}.
\end{equation}
Next we consider the boundary unitarity condition:
\begin{equation}
\sum_{\bar c = \pm} R_a^{\bar c} (\beta) R_{\bar c}^b ( - \beta) =
\delta_a^b.
\end{equation}
Using (26), we get:
\begin{equation}
R_{RL} (\beta) R_{LR} ( - \beta) = 1.
\end{equation}
Note that we have been using the quantity $R_{LR}$ which would
correspond to a left-moving particle being reflected into a right-moving
one. This does not seem to make much sense given that our system is
defined on the half line $(- \infty, 0]$. However, as we will see, it
will prove useful to ignore this and see it only as a formal tool to
derive consistency equations for the boundary reflection factors.

Substituting eq. (28) into (30), we get:
\begin{equation}
R_{RL} (\beta) R_{RL} (i \pi + \beta) = - U_{RL} ( 2 \beta).
\end{equation}
Notice that we cannot take $R_{RL} = R_{LR}$ as can readily be verified if
we substitute $\beta = - i \pi /2$ in eqs.(28) and (30). As we
discussed before if we take the bulk R-L scattering to be trivial
$(U_{RL} \to - \infty)$, then
the exactly screened amplitude (24) of the Kondo problem is a solution
of eq.(31). Let us now consider (31) with nontrivial R-L
scattering. This has the minimal solution:
\begin{equation}
R_{RL} ( \beta) = exp \left(- \frac{i \pi}{4} \right) \left\{
\frac{ sinh \left(\frac{\beta}{2}
    - \frac{i \pi}{8} \right) }{sinh \left( \frac{\beta}{2} + 
    \frac{i \pi}{8} \right) } \right\},
\end{equation}
with $R_{LR} (- \beta) = \left[ R_{RL} (\beta)\right]^*$.
This amplitude has no poles on the physical sheet. The only pole lies
in the second sheet $- \pi < Im \beta < 0$ at $ \beta = - i \pi /4$ and
is associated with the mass scale of the bulk theory. 

Let us now consider the second situation when the boundary has an
effective spin $q=1/2$. This will be denoted as ``{\it free}'' boundary 
condition. We then have:
\begin{equation}
R_a ( \beta) B_b = \sum_{\bar c, d = \pm} R_{a b}^{\bar c d} ( \beta) 
L_{\bar c} ( - \beta) B_d,
\end{equation}
where $B_d$ creates a boundary state with isotopic spin $d= \pm$:
$$
|B>_d = B_d |0>.
$$
The boundary Yang-Baxter equation has to be slightly modified to
incorporate this additional structure, \cite{Ghoshal}:
\begin{equation}
\begin{array}{c}
R_{bc}^{\bar b' c'} ( \beta_2) U_{a \bar b'}^{a' \bar b''} ( \beta_1 +
\beta_2) R_{a' c'}^{\bar a' d} ( \beta_1) S_{\bar b'' \bar a'}^{\bar b
  \bar a} ( \beta_1 - \beta_2) = \\
\\
= S_{a b}^{a' b'} ( \beta_1 - \beta_2 ) R_{a' c}^{\bar a' c'} (
\beta_1) U_{b' \bar a'}^{b'' \bar a} ( \beta_1 + \beta_2 ) R_{b''
  c'}^{\bar b d} ( \beta_2).
\end{array}
\end{equation}
Substituting (14), we get:
\begin{equation}
R_{bc}^{\bar b' c'} ( \beta_2) R_{a c'}^{\bar a' d} (\beta_1) S_{\bar
  b' \bar a'}^{\bar b \bar a} (\beta_1 - \beta_2) = S_{ab}^{a' b'}
  (\beta_1 - \beta_2) R_{a' c}^{\bar a c'} (\beta_1) R_{b' c'}^{\bar b
  d} ( \beta_2).
\end{equation}
We see that the bulk R-L scattering decouples as before. We will see
that it only plays a role in the boundary crossing-unitarity
condition. This, of course, is a consequence of the R-L scattering
being diagonal. Substituting (9), this yields:
\begin{equation}
\begin{array}{c}
\left[ R_{a c'}^{\bar a d} (\beta_1) R_{bc}^{\bar b c'} ( \beta_2) -
  R_{ac}^{\bar a c'} ( \beta_1) R_{bc'}^{\bar b d} ( \beta_2) \right]
  \sigma_T (\beta_1 - \beta_2) = \\
\\
= \left[ R_{bc}^{\bar a c'} (\beta_1) R_{a c'}^{\bar b d} ( \beta_2) -
  R_{a c'}^{\bar b d} (\beta_1) R_{bc}^{\bar a c'} ( \beta_2) \right]
  \sigma_R ( \beta_1 - \beta_2).
\end{array}
\end{equation}
Since the total isospin has to be conserved, we assume the following
$SU(2)$ symmetric combination:
\begin{equation}
R_{ab}^{\bar c d} ( \beta) = \delta_a^c \delta_b^d f_{RL} (\beta) + 
\delta_a^d \delta_b^c g_{RL} ( \beta),
\end{equation}
We then get, using (11):
\begin{equation}
\frac{i}{\pi} ( \beta_1 - \beta_2) g_{RL} ( \beta_1) g_{RL} (\beta_2) =
f_{RL} ( \beta_1) g_{RL} ( \beta_2) - g_{RL} ( \beta_1) f_{RL} (
\beta_2),
\end{equation}
which implies:
\begin{equation}
f_{RL} (\beta) = \frac{i}{\pi} \beta g_{RL} ( \beta).
\end{equation}
The boundary unitarity condition,
\begin{equation}
\sum_{\bar a', b' = \pm} R_{ab}^{\bar a' b'} ( \beta) R_{\bar a'
  b'}^{cd} ( - \beta) = \delta_a^c \delta_b^d,
\end{equation}
reads:
\begin{equation}
\begin{array}{l}
f_{RL} ( \beta) f_{LR} ( - \beta) + g_{RL} ( \beta) g_{LR} (- \beta) =
1,\\
\\
f_{RL} ( \beta) g_{LR} (- \beta) + g_{RL} (\beta) f_{LR} (- \beta) = 0.
\end{array}
\end{equation}
Finally, we consider the boundary crossing-unitarity condition. We
assume the following generalization:
\begin{equation}
R_{\bar a c}^{bd} ( \beta) = R_{\underline b c}^{\underline{\bar a}
  d} ( i \pi - \beta) U_{RL} ( i \pi - 2 \beta),
\end{equation}
where we used the fact that the R-L scattering is diagonal. 
Following Berg et al. \cite{Berg} (cf. section 1), we define the following 
crossing symmetric matrix:
\begin{equation}
G_{bc}^{ad} ( \beta) \equiv R_{\underline b c}^{\underline{\bar a} d}
(\beta) = \delta_b^a \delta_c^d u_{RL} ( \beta) + \delta_{bc}
\delta^{ad} v_{RL} ( \beta).
\end{equation}
As before the matrix $H_{bc}^{ad} ( \beta) \equiv R_{\underline b c}^{\bar d
  \underline a} (\beta)$ vanishes, because it is associated with the exchange 
of momenta, which is not possible since the boundary particle has to stay at 
rest after the interaction. In terms of $u$ and $v$, (42) reads:
\begin{equation}
f_{LR} (\beta) = - U_{RL} ( 2 \beta) u_{RL} (i \pi - \beta), \qquad g_{LR} 
(\beta) = - U_{RL} (2 \beta) v_{RL} (i \pi - \beta).
\end{equation}
The unitarity conditions,
\begin{equation}
\sum_{a', d'} G_{bc}^{a' d'} (\beta) G_{a' d'}^{ad} ( - \beta) =
\delta_b^a \delta_c^d,
\end{equation}
yield the following equations for $u$ and $v$:
\begin{equation}
\begin{array}{l}
u_{RL} ( \beta) v_{LR} ( - \beta) + v_{RL} ( \beta) u_{LR} ( - \beta) +
2 v_{RL} ( \beta) v_{LR} ( - \beta) = 0,\\
\\
u_{RL} (\beta) u_{LR} (- \beta) = 1.
\end{array}
\end{equation}
Notice that if we choose,
\begin{equation}
f_{RL} ( \beta) = - u_{RL} ( \beta) - v_{RL} ( \beta), \qquad
g_{RL} ( \beta) = v_{RL} ( \beta),
\end{equation}
then eq.(38) is automatically satisfied. 
The boundary Yang-Baxter equation for antiparticles yields:
\begin{equation}
u_{RL} (\beta) = \frac{i}{\pi} \beta v_{RL} (\beta),
\end{equation}
which is perfectly compatible with eq.(39) for the choice (47). Solving this 
whole system is tantamount to finding $g_{RL}, g_{LR}$ such that:
\begin{equation}
\left\{
\begin{array}{l}
g_{RL} (\beta) g_{LR} (- \beta) = \frac{\pi^2}{\pi^2 + \beta^2},\\
\\
g_{LR} (\beta) = - U_{RL} (2 \beta) g_{RL} (i \pi - \beta).
\end{array}
\right.
\end{equation}
Suppose that the R-L scattering becomes trivial $U_{RL} \to -1$. In
that case it is perfectly acceptable to take $g_{RL} = g_{LR} = g$:
$$
g ( \beta) g ( - \beta) = \frac{\pi^2}{ \pi^2 + \beta^2}, \qquad 
g( i \pi - \beta) = g (\beta).
$$
This system is solved by $g(\beta) = \sigma_R ( \beta)$, in agreement
with Fendley.

The system (49) with $U_{RL}$ given by eq.(16) is not consistent for 
$g_{RL} = g_{LR} =g$. Again this is checked immediately for $\beta = 
i \pi /2$. It has the minimal solution,
\begin{equation}
g_{RL} ( \beta) = i R_{RL} ( \beta) \sigma_R ( \beta), 
\end{equation}
where $R_{LR}$ is given by eq.(32) and $g_{RL} (- \beta) = \left[ g_{RL} 
(\beta) \right]^*$.

\section{Conclusions}

Let us restate our results. The $\widehat{SU} (2)_1$ WZW theory was defined 
in an axiomatic fashion by introducing the current algebra and constructing 
the conformal symmetry according to Sugawara's procedure. Cardy's approach 
shows that if we impose the conservation of these two symmetries in the 
presence of a boundary there will be two permissible boundary conditions 
which we denoted as fixed and free. There are two RG trajectories that 
terminate at this theory in the IR limit. One of them - the Kondo theory - 
suggests the interpretation of the two boundary conditions as the exactly 
screened and underscreened situations, respectively. The factorized 
scattering is well known for this theory. The second trajectory represents 
the principal chiral model with scale invariant boundary conditions. The 
symmetries of the model, the IR limiting WZW theory and comparison with the 
Kondo model allowed us to construct the corresponding reflection amplitudes, 
(32), (50). The set of equations thus obtained are also valid if we introduce 
an additional boundary perturbation. However it is not clear whether the 
integrability is still preserved.

We will present some checks for these results elsewhere, \cite{Prata}. So far, 
we have computed the ground state energy ({\it $E$-function}) for the system 
defined on an annulus with fixed boundary conditions on both ends, by the 
technique of boundary TBA\footnote{More details about this method can be 
found in refs.\cite{Fendley2}, \cite{LeClair}, \cite{Dorey}}. We had to use 
the fact that the particles obey an exclusion principle and that the backward 
scattering $B^{dc}_{ab} (\beta)$ in eq.(7) is not allowed. In the IR limit 
the $E$-function selects the scaling dimension $\Delta = 0$ in agreement 
with (21) and (23). The {\it $g$-function}, on the other hand, yields 
different boundary entropies at the two extremes of the RG trajectory.

The free boundary condition entails a non-diagonal reflection matrix, thus 
rendering the computations too complicated. The best we can hope for is to 
conjecture the set of TBA equations. 

\vspace{10 mm}

\centerline{\large{\bf Acknowledgements}}

\vspace{10 mm}
The author wishes to thank P.Dorey, E.Corrigan and A.Taormina for useful 
discussions. This work is supported by J.N.I.C.T.'s PRAXIS XXI PhD fellowship 
BD/3922/94.

\bibliographystyle{plain}

\end{document}